\documentclass[floatfix,twocolumn,reprint,showpacs,amsmath,amssymb,xcolor=dvipsnames,pra,superscriptaddress, nofootinbib,longbibliography]{revtex4-2}
\pdfoutput=1

\usepackage[pdftex]{graphicx}%
\usepackage{bm}%
\usepackage[pdftex,colorlinks=true,citecolor=blue,urlcolor=blue,linkcolor=black]{hyperref}
\usepackage{braket}
\usepackage[dvipsnames]{xcolor}
\usepackage[version=3]{mhchem}
\usepackage{float}

\DeclareMathOperator{\e}{e} %
\newcommand{\iu}{{i\mkern1mu}} %
\newcommand{\Ut}{\hat U^{(t)}}
\newcommand{\Uint}{\hat U^{(int)}}
\DeclareMathOperator{\arctantwo}{arctan2}

\begin{document}

\title{Simulating Chemistry with Fermionic Optical Superlattices}

\author{Fotios Gkritsis}
\affiliation{
Covestro Deutschland AG, Leverkusen 51373, Germany
}
\author{Daniel Dux}
\affiliation{
Max Planck Institute of Quantum Optics, Hans-Kopfermann-Str.1, Garching D-85748, Germany
}
\affiliation{Current address: Physikalisches Institut der Universität Heidelberg,
Im Neuenheimer Feld 226, 69120 Heidelberg, Germany}
\author{Jin Zhang}
\affiliation{
Max Planck Institute of Quantum Optics, Hans-Kopfermann-Str.1, Garching D-85748, Germany
}
\author{Naman Jain}
\affiliation{
Max Planck Institute of Quantum Optics, Hans-Kopfermann-Str.1, Garching D-85748, Germany
}

\author{Christian Gogolin}
\affiliation{
Covestro Deutschland AG, Leverkusen 51373, Germany
}

\author{Philipp M. Preiss}
\email{philipp.preiss@mpq.mpg.de}
\affiliation{
Max Planck Institute of Quantum Optics, Hans-Kopfermann-Str.1, Garching D-85748, Germany
}
\affiliation{Munich Center for Quantum Science and Technology (MCQST), Schellingstr. 4, D-80799 München, Germany}

\date{\today}

\begin{abstract}
We show that quantum number preserving Ans\"{a}tze for variational optimization in quantum chemistry find an elegant mapping to ultracold fermions in optical superlattices.
Using native Hubbard dynamics, trial ground states of molecular Hamiltonians can be prepared and their molecular energies measured in the lattice.
The scheme requires local control over interactions and chemical potentials and global control over tunneling dynamics, but foregoes the need for optical tweezers, shuttling operations, or long-range interactions.
We describe a complete compilation pipeline from the molecular Hamiltonian to the sequence of lattice operations, thus providing a concrete link between quantum simulation and chemistry.
Our work enables the application of recent quantum algorithmic techniques, such as Double Factorization and quantum Tailored Coupled Cluster, to present-day fermionic optical lattice systems with significant improvements in the required number of experimental repetitions.
We provide detailed quantum resource estimates for small non-trivial hardware experiments.
\end{abstract}
\maketitle

\section{Introduction}

Finding the ground states of systems of many interacting fermionic particles is one of the central challenges of quantum information science due to their outstanding importance for material science and quantum chemistry.
Even for modest system sizes of several tens of particles, the ability to exactly solve such electronic structure problems can lead to significant real-world advances \cite{mcardle_quantum_2020}.

A promising experimental approach to this challenge is to map many-electron problems to the dynamics of ultracold fermionic atoms in optical lattices \cite{arguello-luengo_analogue_2019, Gonzalez-Cuadra2023}.
Unlike quantum devices operating on spin degrees of freedom, fermionic quantum gases intrinsically respect the exchange antisymmetry of fermionic problems as well as conservation laws for particle number, magnetization, and total spin.
They currently provide the only intrinsically fermionic quantum system in which unitary dynamics can be controlled in real time.
Further, many-body wave functions can be probed on the single-particle level in quantum gas microscopy \cite{Gross2021}.
Exploiting this control over intrinsically fermionic particles broadly for real-world applications is one of the most promising candidates for a useful quantum advantage \cite{Daley2022}.

At present, ultracold fermions in optical lattices are typically used to realize translation-invariant and short-range interacting Hubbard Hamiltonians for specific quantum simulation tasks \cite{Schaefer2020}. Significant algorithmic advances are required to connect these experimental resources to arbitrary many-electron problems without translation symmetry and with long-range interactions as typically encountered in quantum chemistry.

\begin{figure}[H]
    \centering
    \includegraphics[width=\linewidth]{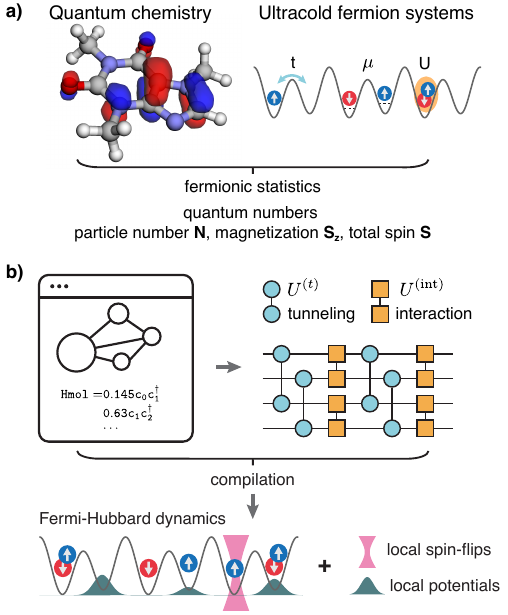}
    \caption{ \textbf{Simulating chemistry with ultracold fermions.} \textbf{a)} A common task in quantum chemistry is to find ground state energies of Coulomb-interacting electrons on molecular structures, usually for fixed particle number $N$, total magnetization $S_z$, and total spin $S$. Synthetic quantum systems with the same fermionic exchange statistics and conserved quantities can be realized with ultracold atoms in optical lattices. \textbf{b)} Using our Ansatz for quantum chemistry problems on optical lattice systems, we prepare molecular trial wave functions via fermionic circuits that can be implemented using Hubbard dynamics and local control of potentials and interaction.}
    \label{fig:qchem_to_optical_lattices_concept}    
\end{figure}

In this work, we show that variational wave functions for molecular Hamiltonians  can be realized in surprisingly simple ways with the dynamics of ultracold fermions in optical lattices. Our Ansatz is inspired by a class of entangling circuits called quantum-number preserving (QNP) fabrics \cite{anselmetti2021qnp}
that were originally developed to represent fermionic wave functions with definite particle number, magnetization, and total spin on qubit-based quantum computers \cite{anselmetti2021qnp}.
These fabrics consist of alternating blocks of gates acting locally on quadruplets of fermionic modes and provide highly expressive Ans\"{a}tze for variational quantum eigensolvers.
By leveraging the mapping between four-fermion terms and Hubbard-like dynamics \cite{Gonzalez-Cuadra2023}, we show that QNP fabrics can be implemented with realistic experimental means in optical superlattices.
We demonstrate that approximate ground states of arbitrary Hamiltonians from quantum chemistry can be implemented as rather simple gate sequences composed of local interaction, chemical potential, and global tunneling operations. Tunable optical superlattices naturally realize the quadruplet structure of fermionic modes required for the fabric and enable block-alternating layers of gates through adjustable dimerization. 

This mapping, in combination with a technique called Double Factorization (DF) \cite{parrish2019quantum,loaiza2022reducing,yen2021cartan, oumarou2022df}, further enables the efficient measurement of quantum chemistry Hamiltonians on fermionic quantum simulators. More generally, the mapping conceptually connects optical lattice systems to the extensive resources developed for spin-based quantum computing and enables the integration of optical lattice simulators into the quantum chemistry workflow. 

The overheads incurred when simulating fermionic systems with qubit architectures due to fermion-to-qubit mappings, have sparked interest in the field of fermionic simulators~\cite{fermionic_mapping_vqe}. To assess the feasibility of our approach, we provide specific examples for the circuits and measurement budgets to perform VQE on small but interesting test molecules with ultracold fermions and numerically explore the precision required on the fundamental tunneling and interaction gates. We find that the required precision and the measurement counts to reduce shot noise to chemical accuracy are challenging but within reach of current experimental hardware \cite{Imperto2023, zhang_superlattice_2023, chalopin_optical_2024}.

\section{Chemistry Hamiltonians and Gate Fabrics}

\begin{figure*}[tb]
    \centering
    \includegraphics[width=\textwidth]{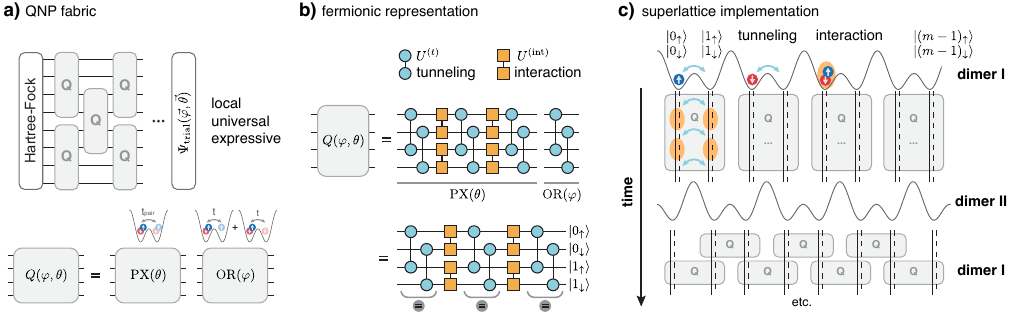}
    \caption{\textbf{Fermion lattice mapping.} \textbf{a)} A QNP (Quantum-Number Preserving) fabric uses a regular brick-layer like structure of four-mode, two-parameter $Q$-gates to form an Ansatz for fermionic wave functions of fixed particle number, magnetization, and total spin.
    Each $Q$-gate element consists of a pair exchange (PX) and orbital rotation (OR) gate. \textbf{b)} Using the decomposition provided in \cite{Gonzalez-Cuadra2023}, we can realize the $Q$-gate as a sequence of tunneling and interaction gates acting between fermionic modes.
    We choose a spin-interleaved ordering of the modes labeled as $|p_{\uparrow/\downarrow}\rangle$ with spatial index $p$ and spin index $\uparrow / \downarrow$. With this choice of mode ordering, tunneling gates occur only between neighboring spatial orbitals of the same spin and interaction gates only between different spin modes within the same spatial orbital.
    Upon compilation, the $Q$-gate has a depth of 5 in the tunneling and interaction gates.
    \textbf{c)} Optical superlattices provide a very natural implementation of the compiled $Q$-gate.
    Each double-well hosts two spatial orbitals with two spin modes each.
    A sequence of spin-independent tunneling and s-wave contact interactions realizes the $Q$-gate on each pair of sites. Changing the superlattice phase in time alternates between different dimerizations, which naturally realizes the QNP fabric.}
    \label{fig:fabric_to_lattice_mapping}    
\end{figure*}

The simulation of chemical reactions is deemed to be one of the most promising early applications of quantum computing.
In order to apply variational quantum algorithms to this task, one needs to have a way of measuring expectation values of the electronic structure Hamiltonian in the state prepared on the quantum device.
In notation that is prevalent throughout quantum chemistry literature, the Hamiltonian has the form of
\begin{align}
\hat{H}
=
E_{\rm c} 
& +
\sum_{pq}^{m-1}
h_{pq}
\hat{E}_{pq} \nonumber \\
& +
\frac{1}{2}
\sum_{pqrs}^{m-1}
(pq|rs)
\left(
\hat{E}_{pq}
\hat{E}_{rs}
-
\delta_{qr}
\hat{E}_{ps}
\right),
\label{eq:2nd_quantized_h}
\end{align}
where $E_c$ is a constant energy offset, $\hat{E}_{pq} = \hat c_{p\uparrow}^\dagger \hat c_{q\uparrow} + \hat c_{p\downarrow}^\dagger \hat c_{q\downarrow}$ is the singlet excitation operator, and $h_{pq}$ and $(pq|rs)$
are the two- and four-index one- and two-body electron integral tensors respectively, whose entries can be efficiently computed classically upon the construction of suitable molecular orbitals.
We will use $p,q,r,s \in \{0,\dots,m-1\}$ to enumerate spatial orbitals and refer to spin orbitals either with tuples such as $p,\uparrow$ and $q,\downarrow$ or indices $i,j,k,l \in \{0,\dots,2m-1\}$ with the convention that $i$ even/odd corresponds to the $p=\lfloor i/2\rfloor$-th spin up/down orbital.
Usually, the Hamiltonian is constructed only on a subset of the $m$ most relevant spatial orbitals, the so-called active space. This implies a challenging scaling of the number of entries in the $(pq|rs)$ tensor as $m^4$. Classically exactly solvable by means of diagonalization are up to about $m=20$ spatial orbitals, highly accurate results can be obtained with reasonable effort by means of Density Matrix Renormalization Group (DMRG) up to about $m=50$, and about $m=100$ is the uppermost limit that can be reached with decreasing accuracy \cite{dmrg_block2}.
This means that only about $100$ sites are needed to push into a regime with potential quantum advantage. 

The challenge posed by the large number of contributions to the energy is exacerbated in conventional qubit based quantum computers by the necessity to map the fermions to the qubit Hilbert space, which turns local fermionic operators such as $\hat c_{\uparrow,p}^\dagger \hat c_{\uparrow,q}$ into non-local qubit operators \cite{mcardle_quantum_2020, Gonzalez-Cuadra2023}. However, even on natively fermionic platforms there is usually no direct way to measure operators such as $\hat{E}_{pq} \hat{E}_{rs}$.
We show how Double Factorization solves this issue in Section~\ref{sec:df}.

When simulating fermionic systems from chemistry, statistical physics, or condensed matter, one is usually interested in obtaining states with given quantum numbers, such as a fixed number of spin up/down electrons and global spin, and when the Hamiltonian is real, its eigenstates are also manifestly real.
In such situations, the ability to prepare states from the respective quantum number sector in a targeted way is desirable. This can offer additional advantages, such as the ability to use postselection on the correct particle number for error mitigation.

These considerations motivated us to explore the use of circuit fabrics of real four-mode gate elements $Q$, each preserving the spin up/down particle numbers and global spin, as shown in Figure~\ref{fig:fabric_to_lattice_mapping}~a). The gate elements $Q$ consist of a fermion pair exchange and an orbital rotation gate, for the preparation of fermionic ground states \cite{anselmetti2021qnp}. A quantum number-preserving gate fabric can be constructed through a block-alternating brick-layer like circuit.
It was demonstrated numerically that at exponential depth this Ansatz is universal, in the sense that every real wave function in a given quantum number sector can be reached and that the Ansatz is highly expressive already at low depths \cite{anselmetti2021qnp}.
For the simplest case of one spin up and one spin down fermion in four orbitals just two parameters are sufficient to reach every molecular ground state.
This approach was further combined with global optimization techniques in \cite{Burton2023} and found to yield highly efficient state preparation circuits. Further, \cite{burton2024accurate} found that a slight re-parametrization of the quantum number preserving gate element can be used to improve trainability.

However, the decompositions into standard qubit gate sets of the quantum number preserving gate element $Q$ described in \cite{anselmetti2021qnp} are system size independent, but relatively deep.
Furthermore, imprecisions in the qubit gates the $Q$ gate element is composed of, usually break the exact quantum number preservation.
Much more compact decompositions into native operations of a proposed neutral ultracold atom platform of essentially equivalent\footnote{A subtle difference is that two different ways of enumerating the fermionic modes were used in the two works; while \cite{anselmetti2021qnp} works in a first all up then all down orbital ordering, \cite{Gonzalez-Cuadra2023} uses an interleaved up, down, up, down, \dots ordering.} gates were recently described in \cite{Gonzalez-Cuadra2023}. The $\hat U^{(pt)}$ gate from this work corresponds to the $\mathrm{QNP}_\mathrm{PX}$ gate from \cite{anselmetti2021qnp} and two of the $\Ut$ gates realize an orbital rotation gate consisting of two Givens gates $G$ as defined in \cite{anselmetti2021qnp}. A definition for both the $\hat U^{(pt)}$ and the orbital rotation gate are given in Appendix~\ref{app:or_and_pt_gate}.

\section{Implementing Quantum Number Preserving Fabrics with Optical Lattices}

\begin{figure*}[t]
    \centering
	\includegraphics[width=\textwidth]{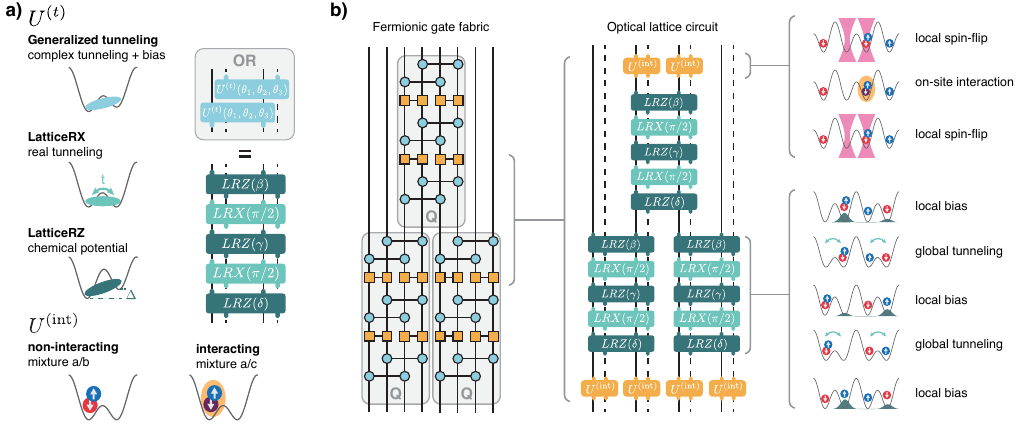}
    \caption{ \textbf{Compilation of circuits to lattice operations.}
    \textbf{a)} The generalized tunneling gate $\Ut$  can be broken down into more elementary lattice operations $\hat{LRX}$ and $\hat{LRZ}$, requiring real-valued tunneling and chemical potential only.
    The $Z3X2$ decomposition requires tunneling only with a fixed phase angle and three chemical potential pulses (see Appendix~\ref{app:zxz_decompositions} for a more compact $ZXZ$ decomposition without this feature).
    The $\Uint$ gate can be implemented by switching between interacting and non-interacting atomic mixtures.
    \textbf{b)} Example compilation of a part of a variational circuit into lattice operations. The sketches show the required lattice operations consisting of local spin-flips, local potential shifts, and global tunneling, as well as the switch of the connectivity using the alternation in superlattice dimerization.}
    \label{fig:circuit_to_lattice_ops_mapping}    
\end{figure*}

We now show how quantum number-preserving gate fabrics can physically be realized with a two-component spin mixture of ultracold fermions in an optical lattice. We choose to work in an optical superlattice geometry as shown in Figure~\ref{fig:circuit_to_lattice_ops_mapping} \cite{Schaefer2020}. An optical superlattice consists of two superimposed, non-interfering lattices whose lattice constants differ by a factor of two \cite{Zhang2023, Bourgund2023}. By independently tuning the depths of the long and short lattices as well as their relative spatial phase, different lattice geometries can be realized. We are especially interested in the regime of strong dimerization, where the lattice depths are chosen such that chains of isolated double-wells with strong intra-well, but negligible inter-well tunneling can be realized. This configuration physically reflects the structure of the QNP fabric, as it supports coherent dynamics within each double-well while completely isolating neighboring double-wells from each other \cite{Imperto2023}. 

In the superlattice, chains of isolated double-wells each supporting a total of four fermionic modes in two spin modes on two sites can be realized. The spin-independent tunnel coupling $t$ within each well can be controlled in real time via the lattice depths. Hubbard $U$ interactions arise from $s$-wave contact interactions between fermions of opposite spins on the same site, but are negligible between different spatial modes.
A spin-independent chemical potential $\mu$ can be applied simultaneously for all double-wells by tuning the relative spatial phase between the long and short superlattices, or on individual sites by microscopically projecting optical chemical potentials \cite{Imperto2023}. Compared to re-configurable tweezer arrays, optical lattices offer larger and significantly more homogeneous systems, at the cost of limited connectivity through nearest-neighbor tunneling only. In the absence of terms with explicit spin dependence, lattice dynamics are $SU(2)$ symmetric and preserve the total spin quantum number $S$.

Our goal is to realize VQE wave functions by a specific time-modulation of the Hubbard parameters $t$, $U$, and $\mu$.
A step towards this goal is the recent demonstration \cite{Gonzalez-Cuadra2023} that fermionic pair tunneling can be decomposed into a sequence of gates consisting only of interaction (Hubbard $U$) and generalized tunneling (Hubbard $t$ and $\mu$) gates, referred to as $\Uint$ and $\Ut$, respectively \cite{Gonzalez-Cuadra2023}.
In this work, we adopt the same definition of gates and visual representation for clarity.
The fermionic representation of the $\Ut$ gate is:
\begin{align}
    \Ut_{i,j} (\theta_1, \theta_2, \theta_3) = e^{
    -i\left[\frac{\theta_1}{2}
    \left( e^{-i\theta_2}c_i^{\dagger}c_j+
    \text{H.c.}
    \right)+
    \frac{\theta_3}{2}(n_i-n_j)
    \right]
    }
    \label{eq:Utdef}
\end{align}
where $c_i^{\dagger}, c_i$ are fermionic creation and annihilation operators and $n_i, n_j$ number operators, with $i,j$ denoting fermionic modes.
The fermionic representation of the $\Uint$ gate is:
\begin{align}
    \Uint_{i,j}(\theta)=e^{-i\theta n_i n_j}
    \label{eq:Uintdef}
\end{align}

The $Q$ gate of the QNP Ansatz consists of correlated pair tunneling followed by an orbital rotation \cite{anselmetti2021qnp}. As shown in Figure~\ref{fig:fabric_to_lattice_mapping}~c), the decomposition from~\cite{Gonzalez-Cuadra2023} can be used to directly realize an instance of the $Q$ gate with the interleaved application of three generalized tunneling pulses $\Ut$ and two interaction pulses $\Uint$ with the correct phase angles. 

Remarkably, the $Q$-gate decomposed into fermionic $\Uint$ and $\Ut$ gates does not involve arbitrary combinations between modes $i$ and $j$ admitted by Eqs.~\eqref{eq:Utdef} and~\eqref{eq:Uintdef}. Instead, only terms that are intrinsically present in optical lattices are required: On-site interactions between pairs of modes, spin-independent nearest-neighbor tunneling, and spin-independent chemical potentials. With sufficient time-dependent control over Hubbard parameters, a fermionic representation of the unit cell of the QNP Ansatz is experimentally straightforward. 

Moreover, the brick-layer structure of the QNP Ansatz finds a very natural realization in optical superlattices: Within each layer, the double-well structure of the lattice enables the decoupled parallel execution of $Q$-gates acting on four fermionic modes. Between layers, the dimerization of a superlattice can be changed via dynamical control of the relative phase between the underlying lattices, coupling alternating pairs of sites in double-well configurations \cite{Imperto2023, chalopin_optical_2024}. Executing $Q$-gates with alternating dimerization of the lattice thus automatically realizes the brick-layer structure of gates required for the QNP fabric.

The remarkably simple mapping of the QNP fabric to the physical optical superlattice is the core observation of this work. Expressed in interaction and generalized tunneling gates, a single realization of the entire $Q$ gate element only requires depth five (see Figure~\ref{fig:fabric_to_lattice_mapping} b)), approximately five times shorter than the decomposition into a standard universal qubit gates set from \cite{anselmetti2021qnp}.
We describe deeper, but experimentally more amenable decompositions in the next section.

\section{Circuit Decomposition for Optical Lattices} 
The fermionic formulation of the QNP Ansatz prepares molecular wave functions on any general fermionic device. Its implementation on specific hardware can be simplified further by transformations to native gate sequences, which we describe for optical lattices below.
The proposed fermionic VQE circuit alternates between terms requiring only single-particle non-interacting tunneling dynamics for $\Ut$ and only two-particle interactions of localized particles for $\Uint$. This structure can be realized by working with an atomic species with Feshbach resonances and at least three available spin states, such as Lithium 6, as shown in Figure~\ref{fig:circuit_to_lattice_ops_mapping}~a) \cite{Schaefer2020}.

We propose to carry out the $\Ut$ gates with a non-interacting spin mixture at a zero crossing of the corresponding scattering length.
The $\Uint$ gates can be performed by locally transferring one of the spin components to a third state, which is interacting with the other component, for example by fast local Raman rotations in a deep lattice. 
The time spent in the interacting mixture before a second spin-flip back to the non-interacting mixture determines the $\Uint$ phase angle, as shown in Figure~\ref{fig:circuit_to_lattice_ops_mapping}\,b). Since the gate is $2 \pi$ periodic, one can always avoid negative gate angles.
As long as the Rabi frequencies for spin rotations are much larger than the on-site interactions $\Uint$, this procedure enables local control over the interaction gate phase angles and isolates the motional from the interaction gates of the sequence.

The tunneling gate $\Ut$ in its original formulation realizes a general orbital rotation requiring \textit{local} control over the chemical potential and the \textit{complex-valued} tunneling in each double-well \cite{Gonzalez-Cuadra2023}. While complex tunneling can be engineered in optical lattices, it is much more desirable to work with the native real-valued tunneling only. We therefore propose to simplify the experimental implementation of $\Ut$ by choosing a decomposition into native single-particle operations consisting of real-valued tunneling and chemical potential only \cite{Imperto2023}, which we refer to as lattice $RX$ ($\hat{LRX}$) and lattice $RZ$ ($\hat{LRZ}$) gates.
\begin{align}
    \hat{LRX}_{p}(\theta) &= \Ut_{2p, 2p+2}(\theta,0,0)\Ut_{2p+1, 2p+3}(\theta,0,0)\\
    \hat{LRZ}_{p}(\theta) &= \Ut_{2p, 2p+2}(0,0,\theta)\Ut_{2p+1, 2p+3}(0,0,\theta)
\end{align}
The arbitrary single-particle unitary $\Ut$ acting in the subspace of two spatial modes can be written for $i \mod 2 = 0$ as a sequence of rotations 
\begin{equation}
    \begin{split}
    \Ut_{i,i+2}(\vec \theta)\Ut_{i+1,i+3}(\vec \theta) = & \hat{LRZ}_{\frac{i}{2}}(\delta) \hat{LRX}_{\frac{i}{2}}(\pi/2)\\
    \hat{LRZ}_{\frac{i}{2}}(\gamma) &\hat{LRX}_{\frac{i}{2}}(\pi/2) \hat{LRZ}_{\frac{i}{2}}(\beta)
\end{split}    
\end{equation}
for appropriate choices of $\beta, \gamma, \delta$ i.e. three chemical potential shifts interleaved with two $\pi/2$ tunneling pulses, as shown in Figure~\ref{fig:circuit_to_lattice_ops_mapping}\,a) and described in Appendix~\ref{app:zxz_decompositions}. Such gates with $i \mod 4 = 0$ correspond to one dimerization of the superlattice and $i \mod 4 = 2$ to the other.

Crucially, for any set of two $\Ut$ acting on one spatial mode, the tunneling gate $\hat{LRX}$ always appears \textit{with the same argument} of $\pi/2$, corresponding to a balanced ``beamsplitter" operation in the double-well \cite{Islam2015}.
The tunneling gate $\Ut$ can therefore be realized by interleaving three \textit{local} chemical potential rotations with two \textit{global} tunneling pulses.
In the superlattice, local chemical potential pulses in each well can be provided by arrays of tightly focused addressing beams \cite{Imperto2023}. Parallel tunneling operations in all double-wells can be realized by ramps of the superlattice depths. Therefore, the $Z3X2$ decomposition proposed above enables the massively parallel implementation of arbitrary $\Ut$ operations, for which local control is required only on the chemical potential, but global control of the tunneling suffices.
To further simplify the application of $\hat{LRZ}$ local chemical potential gates, we use the periodicity of the gate to shift all parameter values to positive (or negative) values, such that all local addressing can be done at a single wavelength creating repulsive or attractive potentials only.

\section{Measuring chemistry Hamiltonians} \label{sec:df}

\begin{table*}[bt]
    \centering
    \begin{tabular}{l|ccclccrr}
         & molecule & active space & basis & RC-DF & layers & shots & depth & qubit depth \\
         \hline
         \#1 & \ce{H4} tetrahedral & (4e, 4o) & cc-pvdz & $n_l=4$ & $4$ & $1.0 \times 10^{5}$ & 83 & 109 \\
         \#2 & \ce{H4} tetrahedral & (4e, 4o) & sto-3g & $n_l=3$ & $6$ & $1.3 \times 10^{5}$ & 117 & 154\\
         \#3 & \ce{H4} square & (4e, 4o) & sto-3g & $n_l=7$ & $7$ & $4.9 \times 10^{5}$ & 134 & 175 \\
         \#4 & \ce{H4} square & (4e, 4o) & sto-3g & $n_l=7$ + FFF & $7$ & $3.0 \times 10^{5}$ & 134 & 175 \\
         \#5 & \ce{HF} distance 1\AA & (10e, 6o) & sto-3g & $n_l=16$ + FFF & $5$ & $5.5 \times 10^{5}$ & 110 & 141 \\
    \end{tabular}
    \caption{Example molecules proposed in this paper as small but interesting test cases for first experimental demonstrations and quantum resources sufficient for simulation with $10^{-3}$ Hartree precision.
    Layers is the number of $Q$ gate layers in a QNP fabric as displayed in Figure~\ref{fig:fabric_to_lattice_mapping}~a) to prepare a sufficiently accurate trial state. Shots is the total number of samples that, if distributed in the right way over the $n_l$ leafs ($n_l$ also equals the number of distinct quantum circuits that need to be run), yields a sub $10^{-3}$ Hartree mean squared error.
    The molecular geometries and Hamiltonians as well as their factorized forms and the shot distributions are available for download from \cite{zenodo_data}.
    Depth is the circuit depth on the fermionic architecture after compilation to the native $\hat{LRX}$, $\hat{LRZ}$, and $\Uint$ gates (and does not include initialization of the modes as occupied or empty), qubit depth is the circuit depth on a qubit device when compiled to controlled Pauli and arbitrary single qubit gates via the decompositions from \cite{anselmetti2021qnp}.
    Two fermionic modes or qubits are required per spatial orbital in the active space.
    RC-DF with (or without) FFF and shot distribution to minimize the variance was performed as described in \cite{oumarou2022df} with a regularization factor of $10^{-4} (10^{-3})$ and RC-DF optimization was aborted once $\Delta_{pqrs} = 10^{-7}$ was reached.
    The active space Hamiltonians in the given basis set were generated with PySCF \cite{pyscf, pyscf_recent}.
    }
    \label{tab:examples}
\end{table*}
Measuring molecular Hamiltonians, even in natively fermionic hardware, is non-trivial because of their large number of terms. Moreover, when written as in Eq.~\eqref{eq:2nd_quantized_h}, the terms are off-diagonal in the computational basis and individually not even Hermitian, thus not accessible with the density measurements realized in experiments \cite{Gross2021}. Naively estimating all their contributions to the energy one-by-one can require a very large number of experimental repetitions (shots).
Double Factorization \cite{parrish2019quantum,loaiza2022reducing,yen2021cartan}, in particular Regularized Compressed Double Factorization (RC-DF) \cite{oumarou2022df} is a technique to overcome these challenges and compress the $(pq|rs)$ tensor by means of classical optimization and find approximate representations of the Hamiltonian described in Eq.~\eqref{eq:2nd_quantized_h}.
The key idea is that, while the Hamiltonian contains up to $m^4$ on- and off-diagonal four-point correlation functions, suitable basis transformations allow the energy to be estimated with sufficient accuracy from only on-diagonal density-density correlations in at most $\mathcal{O}(m^2)$ distinct bases (with RC-DF usually significantly fewer).
The rotations $U_l$ into these bases can be realized efficiently with linear depth orbital rotation circuits in both qubit and fermionic quantum computers \cite{clements2017_lin_depth_orb_rotations}.
The general form of the resulting compressed approximate Hamiltonians is
\begin{equation}
    \hat{H} \approx \hat{H}_{\mathrm{DF}} = U_0\, \hat p^{(1)}\, U_0^\dagger + \sum_{l=1}^{n_l} U_l\, \hat p_l^{(2)}\, U_l^\dagger
    \label{eqn:hamiltonian_doublefactorized}
\end{equation}
with $\hat p^{(1)}$ and $\hat p_l^{(2)}$ respectively degree one and two polynomials of either single mode particle number operators and $n_l \leq m\,(m+1)/2$ (for more details see \cite{oumarou2022df}).
In addition to the reduction of the number of distinct bases and therefore quantum circuits that need to be executed, and simplifying the initial problem statement to straightforward measurements in the particle number basis, the main advantage of RC-DF is a drastic improvement in the variance of the resulting energy estimator.
This results in very significant reductions of the overall number of repetitions needed to determine the molecular energy up to an error $\epsilon$. 
The shot count can be further reduced by combining RC-DF \cite{oumarou2022df} with a technique called Fluid Fermionic Fragments (FFF) \cite{choi2023fluid}, which is a way of optimizing the placement of certain terms of the Hamiltonian that can be accommodated in either $\hat p^{(1)}$ or one of the $\hat p_l^{(2)}$. These techniques reduce the required shot count already for small molecules by one or two orders of magnitude. For illustration, the shot count for chemical precision in the ground state of \ce{H4} (example \#3 and \#4 in Table~\ref{tab:examples}) with $n_l=10$ in a (4e, 4o) active space (four electrons in the four orbitals of the STO-3G basis set) reduces from over $2.4\times10^7$ when measuring the Pauli operators of the Jordan Wigner mapped second quantized Hamiltonian separately to around $4.9\times 10^5$ for RC-DF, to approximately $3.0\times 10^5$ when combining RC-DF with FFF.
As with all sampling based methods, the required shot count scales like $1/\epsilon^2$.

An alternative approach to estimating second and fourth order correlations is to time evolve the state under a fixed quadratic fermionic Hamiltonian for different times and to estimate the correlations from computational basis measurements after the time evolution \cite{denzler2023learning}. While this method is arguably experimentally easier to implement, we expect it to require significantly more shots than RC-DF (with FFF). 

We briefly discuss the prospects of using two other recently proposed methods to obtain molecular ground state energies through the Auxiliary-Field Quantum Monte Carlo (AFQMC) \cite{huggins_unbiasing_2022} and quantum tailored and externally corrected coupled cluster techniques \cite{scheurer2023tailored} in Appendix~\ref{app:tcc}.

Figure~\ref{fig:h4_tetrahedral_compiled_circuit} shows how Double Factorization is realized in fermionic circuits in practice: After the molecular trial wave function is prepared using a brick-wall lattice of $Q$-gates, a unitary basis rotation $U_l$ corresponding to a particular leaf from Eq.~\eqref{eqn:hamiltonian_doublefactorized} is implemented with a circuit of orbital rotation gates composed of $\hat{LRX}$ and $\hat{LRZ}$ gates only. For the next leaf, only the parameters of the final orbital roations are changed to realize a different basis transformation $U_l$, while the preparation of the trial wave function remains unchanged. The total energy of the trial state is obtained by summing over the contributions from all $n_l$ different leafs. 

\begin{figure*}[h]
    \centering
    \includegraphics[width=0.74\textwidth]{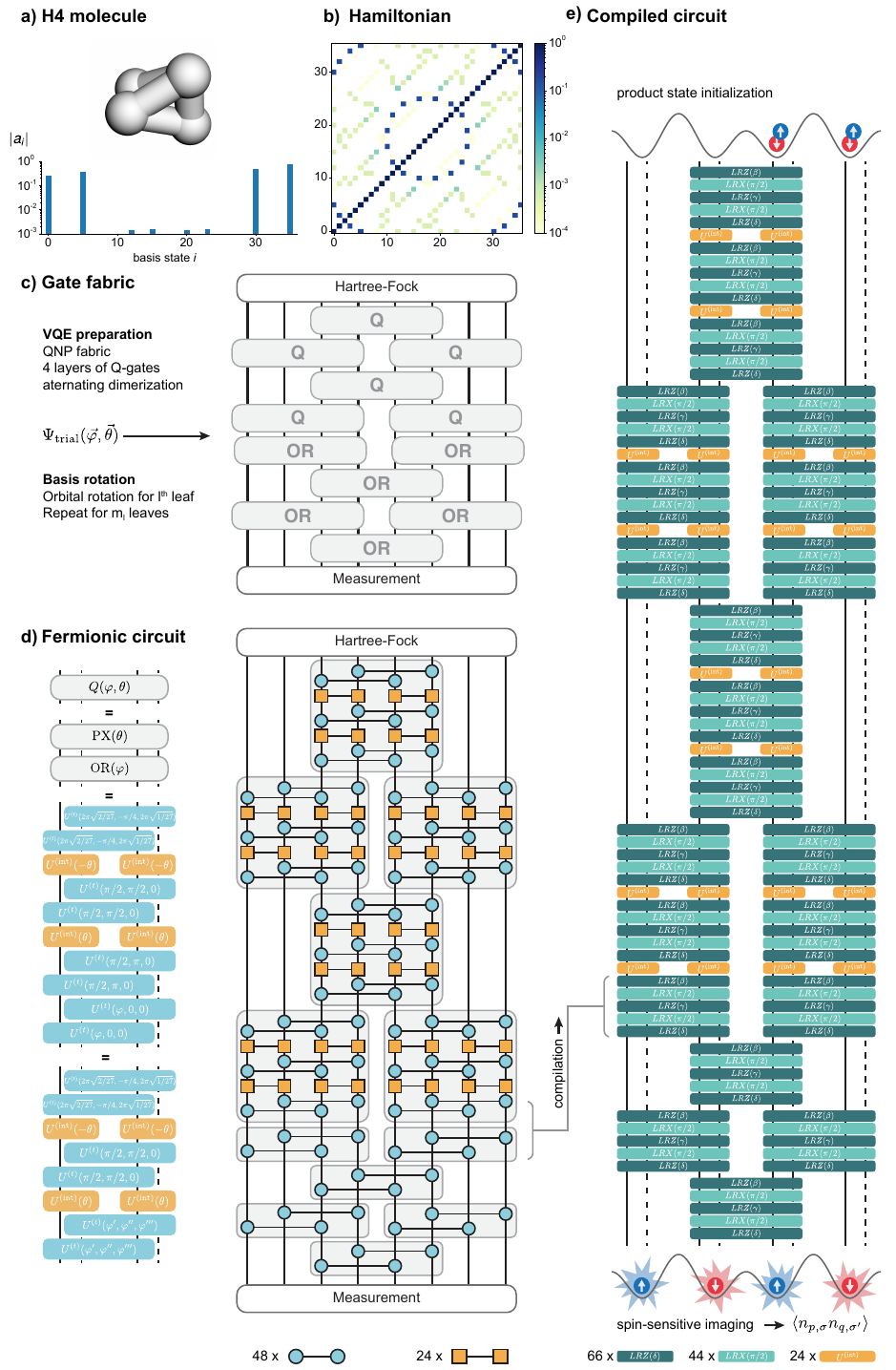}
    \caption{\textbf{Fermionic circuit for the tetrahedral \ce{H4} molecule (example \#1).} \textbf{a)} The tetrahedral \ce{H4} molecule with its ground state wave function represented by its amplitudes $a_i$ in a basis of 36 fermionic states. \textbf{b)} Visual representation of the fermionic Hamiltonian showing the magnitude of its elements normalized to its largest entries, $|H|/\max(|H|)$. \textbf{c)} The gate fabric consists of four layers of $Q$-gates to prepare the trial wave function followed by four layers of orbital rotations. Each leaf of the double-factorized Hamiltonian requires orbital rotations with a different set of angles to be applied to the trial state. \textbf{d)} The full fermionic circuit is obtained by using the decomposition of the $Q$-gate into $\Ut$ and $\Uint$ gates. \textbf{ e)} The full circuit when compiled to the $\hat{LRX}$ and $\hat{LRZ}$ gate set has a depth of 83.}
    \label{fig:h4_tetrahedral_compiled_circuit}
\end{figure*}

\section{Error analysis}

The accuracy with which the ground state energy of a molecular Hamiltonian can be estimated on an optical lattice simulator is limited by statistical shot noise and by the accuracy and precision of the motional and interaction gates performed in the lattice \cite{Dalton2024}.

The motional state of fermions in optical lattices can be affected by several incoherent processes, such as particle loss due to background gas collisions or heating processes to higher bands. Such band excitations effectively introduce distinguishability between fermionic particles and destroy the fermionic exchange statistics in the lowest band. In practice, particle loss rates and band heating rates can be low enough to observe coherent dynamics for hundreds of tunneling times \cite{Gross2021}. We therefore neglect incoherent errors in the following and focus on coherent errors in the unitary gates. We investigate the effect of gate angle errors through a numerical noise model. Here, we are interested in energy errors caused by imprecision in the phase angles of the device-level motional and interaction gates. Such errors spoil the representation of the QNP fabrics in terms of Hubbard dynamics and cannot simply be absorbed in a redefinition of the variational gate angles. Even static miscalibrations will thus not disappear during variational optimization. However, we emphasize that all errors on phase angles on fermionic gates preserve the particle number $N$, the total magnetization $S_z$, and the total spin $S$ and produce valid molecular wave functions. This is in stark contrast to devices that map fermionic problems to spin qubits, where imprecise rotations of individual spins take the wave function out of the correct subspace and produce invalid wave functions that may not be removable by postselection techniques.

To quantitatively assess the impact of phase angle errors, we consider the QNP Ansatz in terms of $\Ut$ and $\Uint$ decomposed into $\Uint$, $\hat{LRX}$, and $\hat{LRZ}$ gates using the $Z3X2$ representation. An example of a full circuit is shown in Figure~\ref{fig:h4_tetrahedral_compiled_circuit}. We then apply a Gaussian distributed, multiplicative noise factor to the angles of the fermionic gates.

We use a `static' noise model to assess the effect of global, constant errors on the phase angles, as they might occur through a global miscalibration of the device's coupling constants. We also consider a `circuit to circuit' noise model, where random multiplicative noise factors are applied to each set of wires individually. These noise factors are kept constant within the computation for a given compiled circuit, but are re-drawn for each new compiled circuit. 
In case of the double-factorized energy measurement there are $n_l + 1$ such circuits, one for each term in Eq.~\eqref{eqn:hamiltonian_doublefactorized}. This model represents the effect of spatially random phase angle errors that slowly drift in time. A detailed description of the two error models is given in Appendix~\ref{sec:error_model}.
The numerical simulations were preformed with PennyLane~\cite{pennylane} with the help of a custom built device using FQE~\cite{fqe_2021} (for more details see Appendix~\ref{sec:auto_diff_and_prameter_shift}) and molecular integrals were computed with PySCF~\cite{pyscf, pyscf_recent}.

\begin{figure}[bt]
    \centering
    \includegraphics[width=\linewidth]{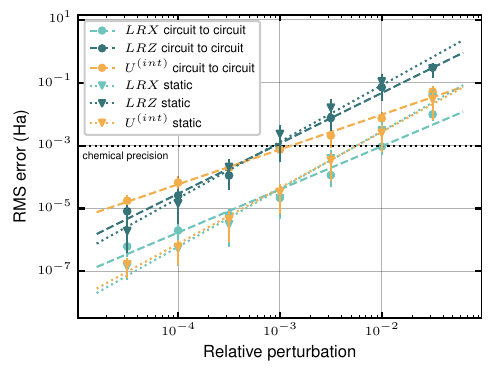}
    \caption{\textbf{Sensitivity to gate errors.} The RMS error of 150 energy calculations of a tetrahedral \ce{H4} molecule (example \#1 in Table~\ref{tab:examples}) without shot noise is plotted for the circuit to circuit noise model and the static noise model applied to one gate type at a time. The dashed and dotted lines are fitted power law functions to improve readability. To reach an error below $10^{-3}$ Hartree in energy, gate imprecisions in the range of $10^{-3}$ to $10^{-2}$ are required.}
    \label{fig:error}
\end{figure}

Figure~\ref{fig:error} shows the inaccuracy of the energy estimate caused by specific gate errors for the different noise models for a distorted tetrahedral \ce{H4} molecule (see table table~\ref{tab:examples},~\#1 and Figure~\ref{fig:h4_tetrahedral_compiled_circuit}). For each simulation, the gate noise is only applied to one type of gates, while the others are assumed to be noise-free, and we compute the root mean square error to the minimal VQE energy of the Ansatz for different realizations of the noise. In the simulated range of perturbations, the RMS energy error scales roughly quadratically with the imprecision of the fermionic gates. Reaching chemical precision ($10^{-3}$ Hartree in energy) requires a relative imprecision of at most $\sim 10^{-3}$ in the phase angle of the $\hat{LRZ}$ and $\Uint$ gates and $\sim 10^{-2}$ in the phase angle of the $\hat{LRX}$ gate in the circuit to circuit noise model. Within the static model, errors have to be below $\sim 10^{-3}$ for the $\hat{LRZ}$ gate and $\sim 7 \times 10^{-3}$ for the $\hat{LRZ}$ and $\Uint$ gates to reach chemical accuracy.

High-quality orbital rotations and entangling gates have already been demonstrated in optical lattices \cite{Islam2015, Zhang2023, Zhu2024}. Recent works have demonstrated local chemical potential rotations \cite{Imperto2023} as well as global tunneling pulses \cite{chalopin_optical_2024} with fidelities of $> 99\%$ averaged over $\sim1000$ sites \cite{Imperto2023}, indicating the potential to execute high-fidelity circuits in parallel on hundreds of fermionic modes.

Besides the precision of the gate operations, a fundamental limit to the measurement of molecular energies is given by the shot noise related to the finite number of measurements. Even with the substantial reduction in shot count requirements that we achieve through the techniques of Regularized Compressed Double Factorization and Fluid Fermionic Fragments, the required sample counts are ambitious for optical lattice simulators (see Table~\ref{tab:examples}). For the concrete example of the tetrahedral \ce{H4} molecule, suppressing shot noise below $10^{-3}$~Ha requires approximately $1.0 \times 10^{5}$ samples from the device. Assuming an experiment cycle time of $\sim 1$~shot/s, a computation time of $\sim 1.2$~days is necessary to compute a single energy at fixed VQE parameters of the tetrahedral \ce{H4} molecule. In superlattices with typical sizes of hundreds of double-wells, several tens of instances of molecular wave functions of the size considered here can be prepared in parallel, reducing the experimental time budget to the scale of hours.

Given these considerations on gate accuracy and shot counts, Table~\ref{tab:examples} suggests use cases for a number of small and artificial but interesting benchmark molecules suitable for first experimental demonstrations. We provide quantum resource estimates (circuit depths and shot counts) for the computation of single point VQE energies at $10^{-3}$ Hartree precision.

\section{Discussion}

In this work we provide a direct mapping that allows the implementation of state-of-the-art variational Ans\"{a}tze from quantum chemistry using fermions in optical superlattices. Remarkably, the ground states of complex molecules can be prepared as gate sequences in almost translationally invariant lattice systems with modest requirements on local control of chemical potentials and local spin flips.

From estimations of the shot budget and simulations of the required stability of the fundamental gates of the lattice system, we anticipate that non-trivial examples of molecular wave functions can be realized on near-future fermionic superlattice simulators and measured in a time budget of several hours. Our approach crucially relies on advanced approximation techniques such as Double Factorization to reduce the number of required shots from impractical to accessible scales.

Our work maintains the known structure of highly expressive QNP fabrics established for quantum chemistry. This choice results in a fixed number of total $Q$-gates in the wavefunction ansatz and a size-independent ratio of circuit depth between fermionic and spin implementations. The fermionic representation of the $Q$-gate has a minimal depth of five and a depth of 17 when compiled to the $Z3X2$ scheme, as compared to 24 when compiled to spin systems (with 14 or 18 two-qubit gates, depending on the exact decomposition) \cite{anselmetti2021qnp}. The chief advantage of the fermionic representation lies in the physical implementation of molecular symmetries, which we expect to become more significant at larger system sizes. Recent, related work \cite{fermionic_mapping_vqe}, has demonstrated a scaling advantage of fermionic over spin simulators for different ansatz choices. 

In principle, optical lattices enable more compact realisations of the fermionic Ansatz than presented here: With optical superlattices implemented along two or three spatial dimensions, $\Uint$ and $\Ut$ gates can be applied along different axes, introducing long-range Q-gates into the tessellation structure of the QNP fabric and further improving its expressiveness. Moreover, direct realizations of locally controlled pair tunneling and orbital rotations could significantly reduce the depth of the compiled fermionic circuits, for example through the Floquet-enhanced pair tunneling already demonstrated in \cite{klemmer2024}. We anticipate that our work will trigger new theoretical and experimental efforts in this direction.

\paragraph*{Acknowledgements}
We thank Fred Jendrzejewski for initiating the project and Torsten Zache and Timon Hilker for helpful conversations. We acknowledge contributions from Oumarou Oumarou, who provided the RC-DF and FFF factorized Hamiltonians. The work was funded by the German Ministry for Education and Research (BMBF) through the projects HFAK (Covestro and MPQ, 13N15630 and 13N16508) and FermiQP (MPQ, 13N15890). The work at MPQ received funding from the European Union’s Horizon 2020 research and innovation program under grant agreement No. 948240 (ERC UniRand).

\paragraph*{Author Contributions}
All authors contributed to the development of the gate decompositions and compilation steps as well as the writing of the manuscript.
PP and DD provided the neutral atom specific know-how, proposed the noise model, and produced the figures.
FG and CG proposed and prepared the test cases and wrote the code enabling the simulations.
Simulation data (Figures~\ref{fig:error} and \ref{fig:EnergyHistogram}) was produced by DD. CG and PMP supervised the project.

\bibliography{fermionicQC}

\clearpage

\appendix

\section{The orbital rotation and pair tunneling gates}
\label{app:or_and_pt_gate}
The $\hat U^{(pt)}$ acts on four fermionic modes and realizes tunneling of a pair of fermions. Its decomposition into $\Ut$ and $\Uint$ gates is adopted from \cite{Gonzalez-Cuadra2023}:
\begin{equation}
\begin{split}
    \hat U^{(pt)}_{i,j,k,l}(\theta) =& e^{-i\left[\theta_1 \left( e^{-i\theta_2}c_i^{\dagger}c_j^{\dagger}c_kc_l+ \text{H.c.}\right) \right]} \\
    =&\Ut_{i,k}(\frac{\pi}{2}, \frac{\theta_2+2\pi}{2}, 0) \Ut_{j,l}(\frac{\pi}{2}, \frac{\theta_2+2\pi}{2}, 0) \\
    &\Uint_{i,j}(\theta_1) \Uint_{k,l}(\theta_1) \\
    &\Ut_{i,k}(\frac{\pi}{2}, \frac{\theta_2+\pi}{2}, 0) \Ut_{j,l}(\frac{\pi}{2}, \frac{\theta_2+\pi}{2}, 0) \\
    &\Uint_{i,j}(-\theta_1) \Uint_{k,l}(-\theta_1) \\
    &\Ut_{i,k}(\frac{2\sqrt{2}\pi}{\sqrt{27}},\frac{2\theta_2-\pi}{4},\frac{2\pi}{\sqrt{27}}) \\
    &\Ut_{j,l}(\frac{2\sqrt{2}\pi}{\sqrt{27}},\frac{2\theta_2-\pi}{4},\frac{2\pi}{\sqrt{27}})
\end{split}
\end{equation}

Two $\Ut$ gates acting on four fermionic modes represent a orbital rotation as presented in \cite{anselmetti2021qnp}:
\begin{equation}
    \hat{OR}_{i,j,k,l}(\theta) = \Ut_{i,k}(\theta) \Ut_{j,l}(\theta)
\end{equation}

\section{The \texorpdfstring{$\Ut$}{g} gate and its decompositions}
A useful way to think about the $\Ut$ gate is that of a single particle tunneling between two spatial modes L and R. Its motional state encodes a single qubit with the computational basis states L and R. A chemical potential between the wells corresponds to a Pauli $\hat Z$ rotation of the qubit. Real-valued tunneling corresponds to $\hat X$ rotations and imaginary tunneling to $\hat Y$ rotations.
With this picture in mind, we can re-write the $\Ut$ gate in terms of logical Pauli operators $\hat X_{i,i+1},\hat Y_{i,i+1},\hat Z_{i,i+1}$.
\begin{equation}
    \Ut_{i,i+1}(\vec{\theta}) = \e^{-\iu (\frac{\theta_1}{2}(\cos(\theta_2) \hat X_{i,i+1} + \sin(\theta_2) \hat Y_{i,i+1}) + \frac{\theta_3}{2} \hat Z_{i,i+1}}
\end{equation}
$\Ut_{i,i+1}$ can thus be understood as a simultaneous rotation through $\theta_1$ about an equatorial axis with azimuth $\theta_2$ and a rotation through $\theta_3$ about the z-axis. For simplicity, the subscripts $i,i+1$ will be suppressed from now on unless necessary.
The generator $A(\theta_1,\theta_2,\theta_3)$ of the $\Ut(\vec{\theta})=\e^{-\iu A(\theta_1,\theta_2,\theta_3)}$ when acting on two sequential fermionic modes and when written as a matrix acting on their Fock space is
\begin{equation}
A(\theta_1, \theta_2,\theta_3) = \frac{1}{2}
\begin{bmatrix}
\label{eq:U_t_qb_form}
    0 & 0 & 0 & 0 \\
    0 & -\theta_3 & \theta_1 k(\theta_2) & 0 \\
    0 & \theta_1 k(\theta_2)^{\dagger} & \theta_3 & 0 \\
    0 & 0 & 0 & 0
\end{bmatrix}
\end{equation}
where $k(\theta_2) = \cos(\theta_2)+\iu \sin(\theta_2)$.
Thus, special cases of $\Ut$ correspond to logical Pauli $X,Y,Z$ rotations
\begin{align}
    \hat X_{i,i+1}(\theta)&=\Ut_{i,i+1}(\theta,0,0)\\
    \hat Y_{i,i+1}(\theta)&=\Ut_{i,i+1}(\theta,\pi/2,0)\\
    \hat Z_{i,i+1}(\theta)&=\Ut_{i,i+1}(0,0,\theta)
\end{align}

\subsection{Decompositions}
\label{app:zxz_decompositions}
The orbital rotation $\Ut$ generally requires access to $\hat X$, $\hat Y$, and $\hat Z$ rotations, i.e. complex-valued tunneling and chemical potential, even if we chose $\theta_2=0$ everywhere. It is useful to decompose $\Ut$ into a sequence containing $\hat X$ and $\hat Z$ rotations only. 

\subsubsection{ZXZ decomposition} 
We can  express $\Ut$ as a $ZXZ$ sequence of rotations, requiring two (simple) chemical potential rotations and a single (more difficult) tunneling/$\hat X$ rotation. Write
\begin{equation}
\label{eq:zxz_decomposition}
    \begin{split}
        \Ut(\vec \theta) &= \hat Z(\delta)\hat X(\gamma)\hat Z(\beta)\\
        &= \Ut(0,0,\delta) \Ut(\gamma,0,0) \Ut (0,0,\beta)    
    \end{split}
\end{equation}

The strategy for obtaining $\beta, \gamma,\delta$ in this scheme lies in making use of the fact that any 2$\times$2 unitary matrix can be decomposed into a sequence of elementary Pauli operations such as $R_Z(\zeta_1)R_Y(\zeta_2)R_Z(\zeta_3)$, where $R_Y, R_Z$ are standard Pauli $\hat Y$ and $\hat Z$ rotations. 
Since the inner 2$\times$2 block of the $\Ut$ gate is also unitary, while the rest of the diagonal elements are equal to one, the task of decomposing $\Ut$ into logical $\hat Z$ and $\hat X$ rotations, becomes equivalent as finding the angles $\zeta_1, \zeta_2, \zeta_3$. Then, it is trivial to incorporate these in the angles $\beta,\gamma,\delta$ in a logical $ZXZ$ setting as:
\begin{align}
    \beta &= \zeta_1 - \frac{\pi}{2} \\
    \gamma &= \zeta_2 \\
    \delta &= \zeta_3 +\frac{\pi}{2}
\end{align}
The code that was employed for performing the decomposition of the $2\times2$ unitary into Pauli $\hat{Y}$ and $\hat{Z}$ rotations was the $\texttt{one\_qubit\_decomposition}$ function found in the PennyLane Python library, v0.35.1 \cite{pennylane}.
\subsubsection{Z3X2 decomposition} 
It is even better to use a decomposition that uses only $\hat Z$-rotations to implement $\Ut$. In that case, only global tunneling ($\hat X$) operations are required and all local operations can be implemented by small local bias fields

\begin{equation}
\label{eq:Z3X2_decomposition}
    \begin{split}
        \Ut(\vec \theta) =& \hat Z(\delta)\hat X(\frac{\pi}{2})\hat Z(\gamma)\hat X(\frac{\pi}{2})\hat Z(\beta) \\
        =& \Ut(0,0,\delta) \Ut(\frac{\pi}{2},0,0) \Ut(0,0,\gamma)\\
        & \Ut(\frac{\pi}{2},0,0) 
        \Ut(0,0,\beta)
    \end{split}
\end{equation}

The methodology for obtaining $\beta, \gamma,\delta$ in this decomposition scheme is very similar to the one in a $ZXZ$ setting, namely, upon finding the $\zeta_1, \zeta_2, \zeta_3$ to construct the inner 2$\times$2 matrix with Pauli $\hat Y$ and $\hat Z$ rotation gates, one can utilize these angles to construct the full $\Ut$ unitary with logical $\hat X$ and $\hat Z$ gates in Eq.~\eqref{eq:Z3X2_decomposition} as:
\begin{align}
    \beta &=\zeta_1 -\pi\\
    \gamma &= -\zeta_2 +\pi\\
    \delta &= \zeta_3
\end{align}

\subsection{Merging of sequential \texorpdfstring{$\Ut$}{g} gates} 
First note that the $\Ut$ gate has the following property when swapping the mode labels it acts on 
\begin{equation}
    \Ut_{i,j}(\theta_1, \theta_2, \theta_3)=\Ut_{j,i}(\theta_1, -\theta_2,-\theta_3) .
\end{equation}
This can be used to make the mode labels of two subsequent gates identical.
Then, in order to construct the generator $A$ of a single $\Ut$ gate that is equivalent to the application of two subsequent $\Ut$ gates with parameters $\vec{\theta}', \vec{\theta}''$, with the structure of $A$ being:
\begin{equation} 
\label{eq:U_t_generator_matrix}
  A = -\iu \log(\Ut(\vec{\theta}') \, \Ut(\vec{\theta}'')) = \begin{bmatrix}
    0 & 0 & 0 & 0 \\
    0 & x & y & 0 \\
    0 & z & w & 0 \\
    0 & 0 & 0 & 0
\end{bmatrix}
\end{equation}
This matrix logarithm is always well defined because its argument is unitary and by taking the (principle) matrix logarithm of $A$, one can find $\Ut$ gate parameters $\theta_1, \theta_2, \theta_3$ that correspond to this generator matrix. The gate parameters $\theta_1$, $\theta_2$, $\theta_3$ of the merged gate can then be inferred from the entries $x, y, z, w$ of the generator.
The $\theta_1$ parameter can be calculated as:
\begin{equation}
    \theta_1=2\sqrt{\left( \frac{|\operatorname{Im}(y-z)|}{2} \right)^2+ \left( \frac{|\operatorname{Re}(y+z)|}{2} \right)^2 }
\end{equation}
The computation of $\theta_2$ is given by:
\begin{equation}
    \theta_2=\arctantwo \left(\frac{\operatorname{Im}(y-z)/2}{\operatorname{Re}(y+z)/2}\right) + \pi
\end{equation}
whereas $\theta_3$ is simply given by: 
\begin{equation}
    \theta_3=\frac{\operatorname{Re}(x-w)}{2}
\end{equation}

\subsection{Autodifferentiability and parameter shift rules}
\label{sec:auto_diff_and_prameter_shift}

It is worth noting that all decomposition and merging steps discussed here and in the main text can be implemented in an auto-differentiable way and the lattice gates $\hat{LRX}$ and $\hat{LRZ}$ satisfy two shift parameter shift rules, which allows the computation of gradients with respect to the parameters $\varphi$ and $\theta$ of the composite QNP gate elements $Q(\varphi, \theta)$ in simulation and on quantum hardware with the help of PennyLane~\cite{pennylane}.
Alternatively, one can use the fact that the $\mathrm{QNP}_\mathrm{PX}$ and $\mathrm{QNP}_\mathrm{OR}$ gates also satisfy multi-shift parameter shift rules~\cite{anselmetti2021qnp} and then compile their parameter-shifted versions.

\section{Molecular energies based on classical shadows}\label{app:tcc}
The recently developed method family of classical shadows  \cite{huang2020predicting,zhao_fermionic_2021,wan2022matchgate,low_classical_2022} opens up an alternative path to estimating molecular energies without a direct measurement of the electronic structure Hamiltonian.
The underlying idea is to repeatedly measure in the computational or particle number basis after appending a randomly drawn circuit from a certain distribution and record for each measurements the results as well as for which random circuit they were obtained.
This data is called the classical shadow of the state before the random circuit.
For certain distributions over random circuits (such as the uniform measure over (passive) fermionic Gaussian  unitaries) it is then possible to efficiently estimate expectation values of certain observables or overlaps with certain classically efficiently describable states (such as computational basis states or Slater determinants).

Particularly relevant in the context of fermionic quantum computing is the construction from \cite{low_classical_2022}, which randomizes over number-preserving (passive) fermionic Gaussian unitaries and allows to compute all overlaps with computational basis states as well as general Slater determinants to error $\epsilon$ from a shadow consisting of just $s \in \mathcal{O}(4\epsilon^{-2}/3)$ shots (which is independent of $m$ and the number of fermions).
Passive fermionic Gaussian unitaries can be realized by means of fabrics of     orbital rotation gates of linear depth, which matches the capabilities of the fermionic neutral atom platform very well.

The overlaps obtained in this way can then be used to guide classical methods for the simulation of electronic structure. This has been demonstrated for Auxiliary-Field Quantum Monte Carlo (AFQMC) \cite{huggins_unbiasing_2022} and split amplitude Coupled Cluster (CC) methods \cite{scheurer2023tailored}.
In the latter case, the resulting energies can be viewed as either an improvement over the standard single reference CC method of computational chemistry, or as a way of augmenting the active space energy of the wave function prepared on the quantum computer with a so-called "dynamic correlation correction" taking care of the energy resulting from the smeared out correlations in the large number of orbitals not in the active space, as well as the coupling between the orbitals inside and outside the active space.
Both methods yield good results from shadows of remarkably low numbers of shots and they have some built-in error mitigation properties \cite{huggins_unbiasing_2022,scheurer2023tailored}, such as robustness to global depolarizing noise.
The latter of the two proposals has the advantage of requiring significantly fewer overlaps and only overlaps with computational basis states \cite{kiser2023classical}, making the post-processing of the results a much less computationally intensive task.

For the overlap estimation protocol from \cite{low_classical_2022} to be applicable to fermionic quantum simulators, the procedure must be modified to use a reference state with the same particle number as the system state and with all particles moved into ancillary modes, instead of the fermionic vacuum state (see Appendix A of \cite{wan2022matchgate} for ways of doing this), since coherent superpositions between fermionic states of distinct particle number are forbidden by particle number superselection rules.
The necessary preparation of the superposition between the system trial state output by a QNP fabric and the reference state can be performed by applying an orbital rotation right after creating the Hartree Fock state, which square-root-swaps the occupied modes with the ancillary modes. The QNP fabric on the system modes then leaves the reference state invariant (owing to it being particle number preserving) and can be applied normally.

\begin{figure}
    \centering
	\includegraphics[width=\linewidth]{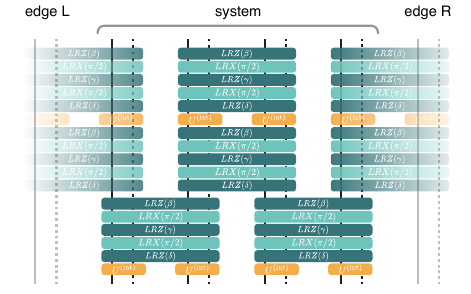}
    \caption{\textbf{System boundaries.} To constrain the fermionic dynamics to a fixed number of modes in the superlattice, the parameters $\beta, \gamma, \delta$ can be chosen such that the corresponding $Z3X2$ block implements the identity. This block is shown shaded in the sequence and prevents any particles from leaking out of the system into the left/right edge modes under the coupling $\hat{LRX}$.}
    \label{fig:system_edges_concept}    
\end{figure}

\section{Error model}
\label{sec:error_model}
To assess the influence of a noisy implementation of the fully decomposed circuit on an optical lattice quantum simulator, we implement a basic noise model. The gate angles of the $\hat{LRX}$, $\hat{LRX}$ and $\Uint$ gates are perturbed by applying a random multiplicative factor that samples a Gaussian distribution, following the specified standard deviation.

In these noise models, we do not model the effect of system boundaries.
Due to the global nature of the $\hat{LRX}$ gate, additional $Z3X2$ blocks have to be appended on the edges of the system that implement the identity, as shown in Figure~\ref{fig:system_edges_concept}.

As the spatial modes on both edges can be prepared without atoms experimentally, errors that lead to the occupation leaking out of the system into the edges can be identified by post- selecting the shots on the atom number in the system.

\subsection{Circuit to circuit noise model}
For the simulation of fluctuations on typical experimental timescales of hours, we implement a noise model that draws a new noise sample for every compilation of a new circuit.
Within the measurement of a given circuit, for the number of individual measurements needed to sufficiently suppress shot noise, this specific realization of noise applied to the circuit is constant.
Further, the same multiplicative factor is applied to the same gate acting on the same set of wires within each circuit.
This noise model therefore models spatially random, but slow drifts of the applied phase angles during computation.

A histogram of the energies calculated when applying this noise model is shown in Figure~\ref{fig:EnergyHistogram}~a).
We attribute the occurrence of energy estimates with energies below the Complete Active Space Configuration Interaction (CASCI) energy of the tetrahedral \ce{H4} molecule to the fact that a new error realization is drawn for every leaf. As the molecular ground state is not the ground state with respect to every leaf, measuring different noisy realizations of the molecular ground state can lead to total energy estimates that are smaller than the actual ground state energy of the molecule.

\subsection{Static noise model}
\begin{figure}
    \centering
	\includegraphics[width=\linewidth]{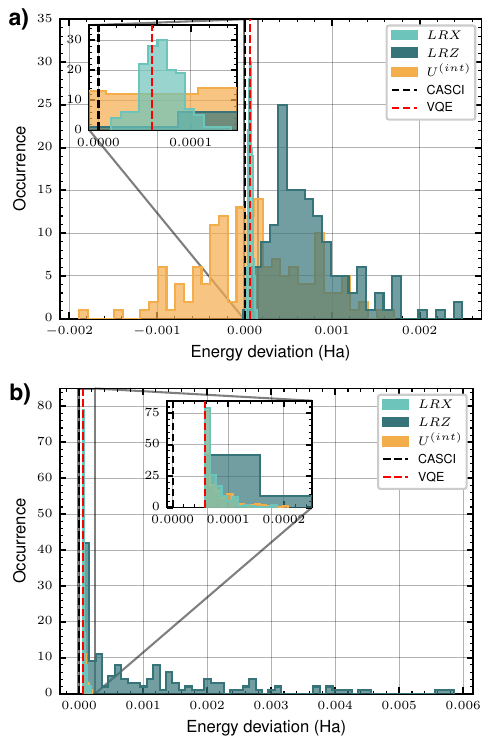}
    \caption{\textbf{Noise model energy histograms.} For the tetrahedral \ce{H4} molecule (example \#1 in Table~\ref{tab:examples}), the histogram displays the energy distribution resulting from a multiplicative error of $10^{-3}$ on the phase angle of one type of gate while assuming the other types of gates are noiseless. We show the distribution for 150 different realizations of the noise. The black dashed line marks the CASCI energy of the molecule, while the dashed red line marks the energy the VQE Ansatz was optimized to. \textbf{a)} The noise model applied is the circuit-to-circuit noise model. \textbf{b)} With the static noise model, only energies larger than the optimized VQE energy are obtained. The increase in energy above the optimal state caused by $\hat{LRZ}$ errors is more than an order of magnitude larger than for $\hat{LRX}$ or $\Uint$ errors, consistent with Figure~\ref{fig:error} in the main text. }
    \label{fig:EnergyHistogram}
\end{figure}

To simulate fully correlated noise, such as a gate miscalibration, we implement a constant multiplicative factor that is applied to all gates of a given type.
The factor is sampled from a Gaussian distribution for every energy calculation.
A histogram of the resulting energies from this noise model is shown in Figure~\ref{fig:EnergyHistogram}~b). 
As we would expect, we only observe energies exceeding the energy the VQE Ansatz was optimized to.

\subsection{Error model for different molecules}
\begin{figure}[H]
    \centering
	\includegraphics[width=\linewidth]{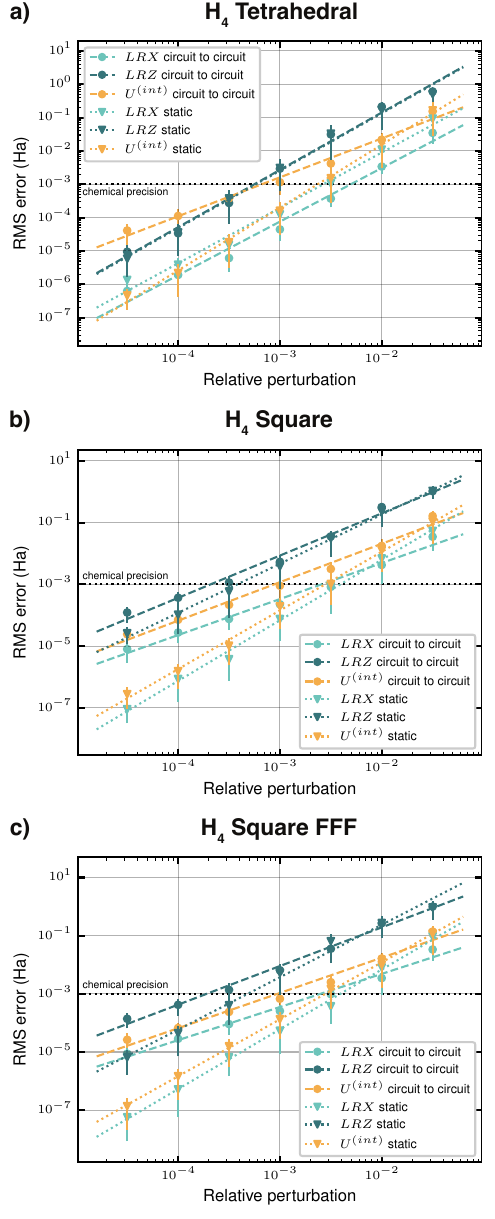}
    \caption{\textbf{Sensitivity to gate errors for different molecules and Ans\"atze.} The RMS error is plotted as a function of the relative perturbation applied to the different gates. The dashed and dotted lines are fitted power law functions to improve readability. In \textbf{a)} the error is plotted for example \#~2 from table~\ref{tab:examples} for 150 calculated energies, \textbf{b)} shows the RMS error of 50 computations for example \#~3 and \textbf{c)} for example \#~4.}
    \label{fig:SupplementErrorModel}
\end{figure}

In Figure~\ref{fig:SupplementErrorModel} we plot the sensitivity of the computed molecular energy to gate errors for different molecules and different Ans\"atze from Table~\ref{tab:examples}. We observe, only minor changes in the scaling of the RMS error with the applied perturbation for a different Ansatz for the same molecule. 
Further, we observe a different scaling behavior between the tetrahedral and the square \ce{H4} molecule.

\end{document}